\documentclass[aps,prd,twocolumn,superscriptaddress,nofootinbib]{revtex4-2}
\usepackage{amsmath}
\usepackage{amssymb}
\usepackage{amsfonts}
\usepackage{amstext}
\usepackage{mathtools}
\usepackage{graphicx}
\usepackage{hyperref}
\usepackage{bm}
\usepackage{microtype}
\usepackage{physics}
\usepackage{xcolor}
\usepackage{tikz}
\usetikzlibrary{patterns,arrows.meta,calc,decorations.pathmorphing}

\hypersetup{colorlinks=true,linkcolor=blue,citecolor=blue,urlcolor=blue}

\begin{document}

\title{Quantum Relaxation Dynamics of Vacuum Energy from Pre-Geometric Gravity}

\author{Andrea Addazi}
\email{andrea.addazi@qq.com}
\affiliation{School of Physics and Astronomy, Anqing Normal University, Anqing 246133, \\
People's Republic of China}
\affiliation{Institute of Astronomy and Astrophysics, Anqing Normal University, Anqing 246133, China}
\affiliation{CAS Key Laboratory for Research in Galaxies and Cosmology, School of Astronomy and Space Science, University of Science and Technology of China, Hefei 230026, China}
\affiliation{Laboratori Nazionali di Frascati INFN, Frascati (Rome), Italy, EU}

\date{\today}

\begin{abstract}
We investigate a quantum relaxation mechanism for the vacuum energy density arising from the spontaneous symmetry breaking (SSB) $SO(1,4)\to SO(1,3)$, driven by a Higgs field $\Phi$ in a MacDowell--Mansouri pre-geometric gravity framework. The theory predicts a countably infinite set of vacua, corresponding to distinct topological sectors, for a periodic Higgs potential with intervening barriers. This structure gives rise to a series of dualities and see-saw relations connecting fundamental scales---the Planck mass and the Cosmological Constant---to topological parameters such as the Gauss-Bonnet coupling and the $\theta$-term, all of which emerge only after SSB. A striking consequence is the quantization of the vacuum energy into discrete levels. Starting from a vacuum expectation value $\langle \Phi\rangle = v_1$ associated with a large vacuum energy, the system evolves via quantum tunneling through the infinite landscape of minima. Crucially, we demonstrate that the transition probability is maximized for the ordered sequence $ \Lambda_1 > \cdots > \Lambda_k > \cdots$, where the Cosmological Constant decreases monotonically toward the asymptotic attractor $\Lambda_\infty = 0$. 
\end{abstract}

\maketitle

\subsection{Introduction}

In a series of recent works, we have elaborated on the idea that General Relativity descends from a spontaneous symmetry breaking (SSB) of an (A)dS isometric gauge group to the Lorentz group, $SO(2,3)$ or $SO(1,4) \to SO(1,3)$ \cite{Addazi:2024rzo,Addazi:2025vbw,Addazi:2025gcl,Addazi:2025zqf,Addazi:2025qkc,Addazi:2026suu,Addazi:2026eto}. This research direction builds upon the foundational work of MacDowell \& Mansouri \cite{MacDowell77} for the kinetic action, as well as subsequent developments by Chamseddine--West \cite{SA1} and Stelle--West \cite{SA2}, and is augmented by a gauge-covariant SSB potential of the type proposed by Wilczek \cite{Wilczek98}.

In this theory, both the Cosmological Constant (CC) and the Planck scale emerge from a recombination of the underlying gauge couplings and the Higgs vacuum expectation value, rather than being fundamental inputs. A natural question is whether such a framework can offer a fresh perspective on the CC problem.
The CC problem represents the most severe Naturalness crisis in modern physics. Standard effective field theory estimates place the vacuum energy at $\rho_{\text{vac}} \sim m_{\text{EW}}^4$ when integrated up to the electroweak scale, dwarfing the observed dark energy $\rho_\Lambda$ by $\sim 10^{60}$ orders of magnitude. If the cutoff is promoted to the Planck scale $M_{Pl}$, this tension escalates to a staggering $10^{120}$~\cite{Weinberg89}. This is not an isolated issue: the electroweak scale itself is destabilized by quadratic radiative corrections that naturally pull $m_{\text{EW}}$ toward the ultraviolet cutoff, demanding an explanation for why the Higgs mass remains so far below $M_{Pl}$.

In this work, we present a realization of quantum vacuum energy relaxation dynamics within the Higgsed phase of MacDowell--Mansouri pre-geometric gravity. Our main result follows from a series of dualities predicted by the theory, which relate the Cosmological Constant, the Planck mass and the gravitational $\theta$-angle. These relations are essential for establishing a foundational principle of the framework: the quantization of the Cosmological Constant into discrete levels, alongside a countably infinite set of vacua for the gravi-Higgs field. These vacua are separated by potential barriers, allowing for quantum tunneling transitions between them. Crucially, we demonstrate that the transition probabilities favor the ordered sequence $\Lambda_1 > \cdots > \Lambda_k > \cdots$, with the preference for smaller $\Lambda$ arising from the increasing number of microstates available at larger $k$. This naturally drives the Cosmological Constant toward smaller values over cosmological time, asymptotically approaching the attractor $\Lambda_\infty = 0$.

\subsubsection{Initial Action}
Our starting point is a gauge theory defined on a bare four-manifold, with no metric or tetrad structure assumed a priori. The gauge group is \(SO(1,4)\), with connection \(A_\mu^{AB}\) and curvature \(F_{\mu\nu}^{AB}\). General covariance is imposed, while the only internal structure is the metric \(\eta_{AB} = \mathrm{diag}(-,+,+,+,+)\) on the tangent space. All action densities are therefore constructed using the topological Levi-Civita symbol \(\epsilon^{\mu\nu\rho\sigma}\).

Let us start with an action composed of three terms: 
\begin{equation}\label{eq:total_action}
\mathcal{S} = \mathcal{S}_{D} + \mathcal{S}_T + \mathcal{S}_{H},
\end{equation}
where $\mathcal{S}_{D} $ is 
the MacDowell-Mansouri Action \cite{MacDowell77}
\begin{equation}\label{eq:MM_action}
\mathcal{S}_{D} = g_{\text{MM}} \int \epsilon_{ABCDE} \, \epsilon^{\mu\nu\rho\sigma} F_{\mu\nu}^{AB} F_{\rho\sigma}^{CD} \Phi^E \, d^4x\, ,
\end{equation}
where \(\Phi^A\) is the Gravi-Higgs field in fundamental representation of \(SO(1,4)\), \(g_{\text{MM}}\) is the MM coupling constant -- mass dimension \([\phi]^{-1}\).

The topological Pontryagin \(\theta\)-term, indicated as $S_T$, reads as 
\begin{equation}\label{eq:theta_action}
\mathcal{S}_T = \frac{\theta}{32\pi^2} \int \text{Tr}(F \wedge F) 
\end{equation}
with $\theta$ an angular parameter. 

The $\mathcal{S}_H$ action reads as 
\begin{equation}\label{eq:periodic_potential}
\mathcal{S}_{H} = \int |\mathcal{J}| \, \bar{\mathcal{V}}(|\Phi|)\, d^4 x\, \, , 
\end{equation}

where
\begin{equation}
\label{JJJ}
\mathcal{J} = \epsilon_{ABCDE}\epsilon^{\mu\nu\rho\sigma}\nabla_\mu\Phi^A\nabla_\nu\Phi^B\nabla_\rho\Phi^C\nabla_\sigma\Phi^D\Phi^E
\end{equation} is the pre-geometric gauge volume form 
and 
$\bar{\mathcal{V}}(|\Phi|)\equiv \bar{\mathcal{V}}(\sqrt{|\eta_{AB}\Phi^A\Phi^B|})$  a generic functional of the scalar field leading it 
to a VeV. 
After SSB, the $\mathcal{J}$-form corresponding to Eq.\ref{JJJ} reconstructs the covariant 4-volume form.

\subsubsection{Higgsed Phase and General Relativity}

When the Higgs gets VeV as
 \(\langle \Phi^A \rangle = v \delta_5^A\), the gauge potential is split into tetrads and spin connection:
\begin{equation}\label{eq:decomposition}
A_\mu^{a5} = \epsilon e_\mu^a, \qquad A_\mu^{ab} = \omega_\mu^{ab},
\end{equation}
with \(e_\mu^a\) the tetrads, \(\omega_\mu^{ab}\) the spin connection, \(a,b = 0,\dots,3\) Lorentz indices
where the scale $\epsilon$ has the same dimensions as $\Phi$, rendering the gauge potentials and tetrads dimensionally consistent.

After the SSB \cite{Addazi:2024rzo}, the MM term will give: 
\begin{equation}\label{eq:SSB_result}
\mathcal{S}_{D} \xrightarrow{\text{SSB}} \int d^4x \, e \left( \frac{M_{\text{P}}^2}{16\pi} R - \frac{M_{\text{P}}^2}{8\pi} \Lambda + \alpha_{\text{GB}} \mathcal{G} \right),
\end{equation}
where \(e = \det e_\mu^a\), \(R\) is the Ricci scalar and \(\mathcal{G}\) the Gauss-Bonnet scalar density.
The initial theory parameters $(g_{\text{MM}}, v,\epsilon)$ are recombined into $(M_{\text{P}},\Lambda,\alpha_{GB})$ as 
\begin{equation}
M_{\text{P}}^2 = 32 g_{\text{MM}} v \epsilon^2,\quad  \Lambda = 3\epsilon^2,\quad \alpha_{\text{GB}} = -4 g_{\text{MM}} v. \label{eq:GB_coupling}
\end{equation}
This also leads to an insightful relation between the three gravitational parameters 
as 
\begin{equation}
\label{Ins}
\alpha_{GB}=-\frac{3M_{Pl}^2}{8\Lambda}\, . 
\end{equation}
We stress that this relation is a pure prediction of MM theory while the three parameters 
in GR are not connected. 



After the SSB, the topological \(\theta\)-term described by \(\mathcal{S}_T\) contains the Holst term  \cite{Holst96}: 
\begin{equation}\label{eq:Ree_term}
\mathcal{S}_T\supset \frac{1}{\gamma} \int \epsilon^{\mu\nu\rho\sigma} R_{\mu\nu}^{ab} e_{\rho a} e_{\sigma b} \, d^4x\, .
\end{equation}
In addition the $Tr(F\wedge F)$ term generates the Pontryagin density, corresponding to $R^{ab}\wedge R_{ab}$,
and a $T_{a}\wedge T^{a}$ involving torsion.
In particular,
\begin{equation}
\label{dec}
\begin{split}
\frac{\theta}{32\pi^{2}}\,\mathrm{Tr}\!\left(F\wedge F\right)
&= -2e\,\frac{\theta}{32\pi^{2}}
\Bigl[\,R^{ab}\wedge R^{ab} \\
&\qquad - \bigl(-2\epsilon^{2}\,T_{a}\wedge T^{a}
+ 4\epsilon^{2}\,R^{ab}\,e_{a}\wedge e_{a}\bigr)\,\Bigr],
\end{split}
\end{equation}
$(e={\rm det} e_{\mu}^{a})$ where the last combination in paranthesis is the Nieh-Yan (NY) density $d(e^{a}\wedge T^{a})$ \cite{NY} multiplying 
a constant $\beta$. 
From this we can identify the Pontryagin angle as
 \begin{equation}
 \label{thetaP}
 \theta_{P}=-2\theta
 \end{equation}
and the last terms have coefficients as 
\begin{equation}
\label{alpha}
\alpha=-\frac{\theta}{4\pi^{2}}\epsilon^{2}=\frac{\theta_{P}}{2\pi^{2}}\epsilon^{2}\, ,
\end{equation}
and 
\begin{equation}
\label{alphabeta}
\beta=-2\alpha=\frac{\theta}{8\pi^{2}}\epsilon^{2}=-\frac{\theta_{P}}{4\pi^{2}}\epsilon^{2}\, ,
\end{equation}
with $\beta=2/\gamma$ and $\gamma$ the Immirzi parameter
 \cite{Immirzi97} parameter that, surprisingly, 
is related to the $\theta$-angle and the CC: 
\begin{equation} 
 \frac{1}{\gamma} = \frac{\theta \Lambda}{12\pi^2 M_{Pl}^{2}}= \frac{\theta}{|\theta_{GB}|}=\left|\frac{\theta_{P}}{2\theta_{GB}}\right|\, ,
  \label{Immirzi}
\end{equation}
with $|\theta|_{GB}\equiv 32\pi^2 |\alpha_{GB}|$ and $\epsilon^{2}=\Lambda/3$.
Moreover, the Immirzi parameter is here also identified as the inverse of the NY parameter. In general, the torsion is not absent,
rendering the emergent Holst term as non-topological. These results correspond to a special case of a generalized pre-metric model analysis recently presented in \cite{Addazi:2026suu}.
The Immirzi parameter is conventionally viewed as a quantum ambiguity, as it does not affect the classical equations of motion while influencing quantum dynamics, particularly in Loop Quantum Gravity (LQG)~\cite{Rovelli2004}. In our framework, this parameter emerges as the ratio of $\theta$-angles, thereby unifying the Immirzi ambiguity with the more familiar $\theta$-ambiguity of gauge theories.

It is noteworthy that for $\gamma=1$, the formalism naturally reproduces results obtained via the Chern-Simons-Kodama state approach, suggesting an emergent gravitational quantum Hall effect~\cite{CSK1,CSK2,Alexander:2022ocp,Magueijo:2020ugp,Alexander:2025qkx}.

\subsection{Topological Sectors and periodic potential}

Eq.\ref{Immirzi} non-trivially related the topological angles to the CC and to the original VeV of the Higgs. 
This is the key observation on the basis of our following arguments. 
As also in Yang-Mills theory, in gravity the angle \(\theta\) is periodic: \(\theta \sim \theta + 2\pi\) leaves the path integral unchanged. 
Less trivial from the YM prospective but true for 4d gravity is that also $\theta_{GB}\equiv 32\pi^2 |\alpha_{GB}|$ 
has the same periodicity, because the GB term is a global integral and a topological invariant. 
Thus, we have a unified periodicity law
\begin{equation}
\theta_{GB}^{(k)}= \gamma \theta^{(k)} = \theta_{GB}^{(1)} + 2\pi k , \qquad k \in \mathbb{Z}\, .
\end{equation}
Using Eq.\ref{Immirzi}, we obtain that the CC is quantized as follows 
\begin{equation}
\Lambda_k = \frac{1}{32\pi^{2}}\frac{3M_{Pl}^2}{8(\theta_{GB}^{(1)} + 2\pi k)}\, ,
\end{equation}
or equivalently the vacuum energy levels 
\begin{equation}
\label{level}
\epsilon_{k}=\frac{1}{16\pi}\frac{M_{Pl}}{\sqrt{\theta_{GB}^{(1)} + 2\pi k}}\, . 
\end{equation}

The zero-level $k=1$ 
corresponds to $\epsilon_{1}=M_{Pl}/16\pi\sqrt{\theta_{GB}^{(1)}}$ (planckian energy level)
while 
$k\sim 10^{120}$ will reconstruct the observed value of $\Lambda$
corresponding to the large k-cycle limit $\epsilon_{k}\sim M_{Pl}/\sqrt{k}$
and $\Lambda \sim M_{Pl}^{2}/k$. 
The Planck Mass $M_{Pl}$ does not change with k
as the gravitational Newton constant remains universal 
for all vacuum states. 
Let us note that the periodicity of the cosmological constant
can also be inferred solely from the periodicity of the Gauss--Bonnet
angle --- without necessarily invoking the proportionality of the
$\theta_{P}$ angle to the Immirzi parameter --- through the direct
relation Eq. (\ref{Ins}), as argued in \cite{Addazi:2026eto}. Indeed, any periodic rotation
of the Gauss--Bonnet angle leaves the state invariant, since it does not
change the path integral. We recall that, in four dimensions, the
integral of the Gauss--Bonnet density is associated with the topological
Euler number.

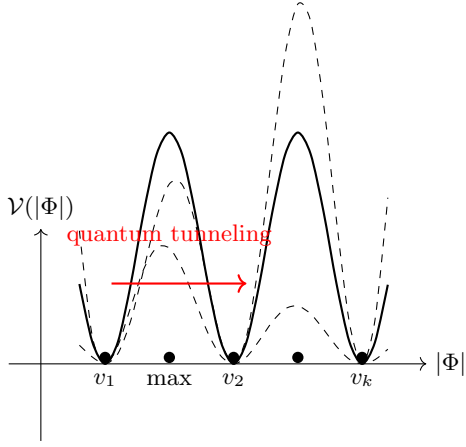
\begin{figure}[t]
\centering
\begin{tikzpicture}[scale=0.85]
\draw[->] (-0.5,0) -- (6,0) node[right] {$|\Phi|$};
\draw[->] (0,-1.2) -- (0,2.10) node[above] {$\mathcal{V}(|\Phi|)$};

\draw[domain=0.6:5.4, smooth, variable=\x, black, thick] 
    plot (\x, {1.8-1.8*cos(deg(3.14*\x-3.14))});
    
    \draw[domain=0.6:5.4, smooth, variable=\x, black, dashed] 
    plot (\x, {0.7*\x-0.7*\x*cos(deg(3.14*\x-3.14))});
    
       \draw[domain=0.6:5.4, smooth, variable=\x, black, dashed] 
    plot (\x, {1.8*(1/\x)-1.8*(1/\x)*cos(deg(3.14*\x-3.14))});

\foreach \x in {1,2,3,4,5}
    \fill (\x,0.1) circle (2.5pt);

\node at (1, -0.25) {$v_1$};
\node at (2, -0.25) {max};
\node at (3, -0.25) {$v_2$};
\node at (5, -0.25) {$v_k$};

\draw[->, red, thick] (1.1,1.25) -- (3.2,1.25);
\node[red] at (2.0, 2.0) {quantum tunneling};

\end{tikzpicture}
\caption{Periodic Higgs potential $\mathcal{V}(|\Phi|)$ from Eq.~\eqref{eq:periodic_potential2}. The solid black curve corresponds to the case where the Gravi-Higgs excitation mass $m_\rho$ is Planckian and $k$-scale invariant, with constant barrier heights between minima. The dashed curves illustrate possible deformations of the potential arising from $k$-dependent scalings of $m_\rho^{(k)}$, which modify the barrier heights and shapes. Multiple degenerate minima are labeled by the topological quantum number $k$, located at $v_1, \dots, v_k$, each corresponding to a different vacuum sector with distinct Cosmological Constant $\Lambda_k$. Quantum tunneling transitions between adjacent vacua are indicated by the red arrow.}
\label{fig:potential}
\end{figure}

The fact that the $\theta$-angle is periodic also immediately 
implies monodromy of the Higgs potential and
and countable infinite number of possible VeV for the 
Higgs as 
\begin{equation}\label{eq:v_quantized}
\langle \Phi \rangle_k=v_k \sim |\theta^{(k)}| \sim k,
\end{equation}
with $(M_{Pl}=1)$. 

 The Higgs VEV is quantized and equal to the same integer that labels the vacuum. The VEV periodicity allows to fix a class of Higgs potential as follows:
\begin{equation}\label{eq:periodic_potential2}
\mathcal{V} = h_{\text{SSB}} |\mathcal{J}| \left[1 - \cos\left(128\pi^2 g_{\text{MM}}(|\Phi| - v)\right)\right]^z,
\end{equation}
where $h_{\text{SSB}}$ is a new constant controlling the potential 
and $g_{\text{MM}}$ is the MM coupling; $z\geq 0$ is a power-law index.
This potential resembles the axion-like case, with several minima separated by large barriers (See Fig.\ref{fig:potential}). In our scenario, each minimum corresponds to a different cosmological constant.
In case ultimately $\theta$ is promoted to a dynamical gravi-axion, its monodromy is related to the Higgs one. 
Note that such shifts are not related to the periodicity of a phase-field. The minima are not gauge-equivalent points of a $U(1)$ target space, but genuinely different vacua labeled by the topological integer $k$: the integer \(k\) labels distinct radial fluctuation fields $\rho$, with 
different masses  \(m_\rho^{(k)} \sim M_{Pl}\,k^{-p}\) and cosmological constants $\Lambda_k\sim M_{Pl}/k$.

Expanding the potential around the minimum $\Phi^5 = v + \rho$, the quadratic mass term for the radial fluctuation $\rho$ scales as
\begin{equation}
\label{murho}
m_{\rho}^2/{M_{Pl}^{2}} \sim h_{SSB}g_{MM}^2 \epsilon^4 v^5,
\end{equation}
reminding that $[h_{SSB}]=M^{-5}$ and $[g_{MM}]=M^{-1}$.

This suggests that the scalaron mass may acquire a $k$-dependence. Substituting the scalings $\epsilon \sim 1/\sqrt{k}$ and $v \sim k$ into Eq.~\eqref{murho}, while leaving the $k$-dependence of the couplings unspecified, yields

\begin{equation}
\label{murho_scaling}
m_\rho^2/M_{Pl}^2 \sim k^{-c_{ssb}} k^{-c_{MM}} k^{3},
\end{equation}

where $c_{ssb}$ and $c_{MM}$ are undetermined exponents characterizing the running of the SSB and MM couplings $h_{SSB,}g_{MM}^2$, respectively.

If $-c_{ssb}-c_{MM}+3>0$, the scalaron mass grows with $k$, becoming super-Planckian for sufficiently large topological sectors. This would bear an intriguing analogy to Kaluza-Klein towers, where mode masses can extend well beyond the Planck scale. A more conservative requirement is $-c_{ssb}-c_{MM}+3 \leq 0$, ensuring $m_\rho^{(k)} \leq M_{Pl}$ for all $k$ (up to order-one numerical coefficients), corresponding to $m_\rho^{(k)} \sim M_{Pl}/k^{c_{ssb}+c_{MM}-3}$.

A lower bound on the scalaron mass follows from the requirement that it cannot be lighter than the Hubble scale, $m_\rho^{(k)} \gtrsim \sqrt{\Lambda_k} \sim M_{Pl}/\sqrt{k}$, otherwise it would behave as an effectively massless field. This imposes $-c_{ssb}-c_{MM}+3 \geq -1/2$. The saturation of this bound, $c_{ssb}+c_{MM} = 7/2$, corresponds to a superlight scalaron that could serve as a field candidate for dynamical dark energy.

The limiting case $c_{ssb}+c_{MM}=3$ is particularly interesting, as it yields $m_\rho^{(k)} \sim M_{Pl}$ for every $k$, implying that the scalaron mass is invariant under $k$-translations. Assuming the mass scaling is predominantly controlled by the SSB parameters rather than the MM kinetic term, a natural choice is $c_{ssb}=-3$ and $c_{MM}=0$, corresponding to a factor $v^{-3}$ in $h_{SSB}$, i.e., $h_{SSB} \sim M_{Pl}^{-2} v^{-3}$. In such a case, the $\rho$-field can provide a natural candidate for super-heavy dark matter.

\subsection{Quantum Tunneling Transitions and Relaxation}
Let us now estimate the quantum tunneling.
The Euclidean bounce action was originally introduced in 
Ref.~\cite{Coleman1977}, and a pedagogical discussion can be 
found in Ref.~\cite{Coleman1985}. Finite-temperature effects 
were treated in Ref.~\cite{Linde1983}.
In thick wall approximation, the Euclidean bound action corresponding to the 
instanton\footnote{We remark that our solutions for the Euclidean
action are approximate, and we do not possess full classical instanton
profiles for the Euclidean field equations, which appear to be
analytically intractable. Nevertheless, the presence of finite barriers
guarantees the existence of quantum tunneling solutions, which are
typically in correspondence with instantons. A more detailed study of
explicit instanton solutions, or their numerical integration, is left
beyond the scope of this work.}
\begin{equation}
\label{Bounce}
\mathcal{S}_E^{(k)} \sim R_{\rho,k}^4 \Delta \mathcal{V}_k/e \sim \Delta \mathcal{V}_k/m_\rho^4 e\, . 
\end{equation}
with $R_{\rho}$ the scalar compton length of the scalar and relevant integration cutoff for the euclidean integral: 
where
\begin{equation}
\label{DeltaV}
\Delta \mathcal{V}_k=\mathcal{V}_{k+1}-\mathcal{V}_{k} \sim h_{SSB}^{(k)}  \epsilon_k^4 v_k^5 e \sim g_{MM(k)}^{-2}m_{\rho(k)}^2\, e \, . 
\end{equation}
Therefore
\begin{equation}
\label{Bounce2}
S_E^{(k)} \sim   g_{MM(k)}^{-2}m_{\rho(k)}^{-2}. 
\end{equation}

In the large-$k$ limit, the thin-wall approximation becomes physically appropriate, where $S_E^{(k)} \sim \sigma_k R_k^3$ with $R_k \sim m_{\rho(k)}^{-1}$. For tension $\sigma_k \sim (g_{MM(k)}^2 R_k)^{-1}$, this yields the holographic scaling $S_E^{(k)} \sim g_{MM(k)}^{-2} R_k^2 \sim g_{MM(k)}^{-2} m_{\rho(k)}^{-2}$, matching the thick-wall result.

\subsubsection{Density of states and Quantization of de Sitter Entropy}

The Gibbons--Hawking formula for the de Sitter entropy \cite{GB} gives
\begin{equation}\label{eq:dS_entropy}
S_{\text{dS}} = \frac{3\pi M_{Pl}^{2}}{\Lambda}.
\end{equation}

Substituting the quantized spectrum \(\Lambda_k \sim M_{Pl}^2/k\) from Eq.~\eqref{level}, we obtain
\begin{equation}\label{eq:dS_quantized}
\boxed{S_{\text{dS}}^{(k)} = \frac{3\pi M_{Pl}^2}{\Lambda_k} \sim k \sim v_k}\, .
\end{equation}

The de Sitter entropy is therefore itself quantized, taking discrete values proportional to the topological index \(k\). Remarkably, the observed value \(S_{\text{dS}} \sim 10^{120}\) is precisely identified with the large topological quantum number labeling our present vacuum sector. Moreover, since \(v_k \sim k\), the entropy is proportional to the vacuum expectation value of the pre-geometric Higgs field, suggesting that \(\Phi\) may be interpreted as an {\it entropy field} whose VEV encodes the information content of the cosmological horizon.

This quantization suggests that the de Sitter horizon can be viewed as a {\it topological quantum memory}, with each unit of entropy corresponding to a topological qubit. This interpretation is supported by the scaling relations
\begin{equation}
|\alpha_{GB,k}| \sim |\theta| \sim v_k \sim S_{\text{dS}} \sim k,
\end{equation}
which establish a direct correspondence among the Gauss-Bonnet coupling, the gravitational \(\theta\)-angle, the Higgs VEV and the de Sitter entropy. Conversely, the gravitational coupling at the energy \(\epsilon_k\) scales as \(|\alpha_{G,k}| \sim G_N \Lambda_k \sim 1/k\), suggesting a duality \(|\alpha_{GB}| \, \alpha_G \sim 1\) that is invariant under \(k\)-transformations.

The total storage capacity of the Universe is then fixed by the topological index \(k\), with the horizon encoding \(k \sim 10^{120}\) qubits of quantum information. In this picture, the Cosmological Constant is not merely an energy density but rather a measure of the information density stored on the horizon: a smaller \(\Lambda\) implies a larger horizon area and, correspondingly, a larger quantum memory capacity. The discrete spectrum \(\Lambda_k \sim 1/k\) thus acquires a direct information-theoretic meaning: each vacuum sector corresponds to a Universe with a different maximum storage capacity and our observed vacuum is one with an enormous information content.
The asymptotic attractor $\Lambda_{\infty}=0$ corresponds to an infinite memory capacity. 

Moreover, the quantization of the Cosmological Constant implies
\begin{equation}\label{eq:decomposition}
e_\mu^{a(k)} = \frac{1}{\epsilon_k} A_\mu^{a5}, \qquad \omega_\mu^{ab} = A_\mu^{ab},
\end{equation}
with $\epsilon_k = \sqrt{\Lambda_k/3}= M_{Pl}/\sqrt{S_{dS,k}/\pi}$. 
This reveals a degeneracy between tetrads and spin connections: while the tetrads acquire a $k$-dependent scaling, the spin connection remains $k$-independent. This is consistent with the metricity condition 
\begin{equation}
\nabla_\mu e_\nu^{a(k)} + \omega_{\mu\ b}^{\ a} e_\nu^{b(k)} = 0,
\end{equation}
where $\nabla_\mu$ is the covariant derivative defined with respect to the spin connection $\omega_\mu^{ab} = A_\mu^{ab}$. The $k$-factors cancel identically, leaving no $k$-dependence for the spin connection.

Since tetrads encode the metric structure—and ultimately gravitons upon quantization—we may rewrite the metric tensor suggestively as
\begin{equation}
\label{suggestive}
g_{\mu\nu}^{(k)} = \eta_{ab} e_\mu^{a(k)} e_\nu^{b(k)} = \pi^{-1}L_{Pl}^2 \, S_{dS,k} \, A_\mu^{a5} A_\nu^{a5},
\end{equation}
exhibiting a direct proportionality between the metric and the de Sitter entropy, scaling as $k$. Equivalently, in terms of the Higgs VEV, $g_{\mu\nu}^{(k)} \sim \langle \Phi \rangle_k A_\mu^{a5} A_\nu^{a5}$. The latter suggests that gravitons dissolve in the limit $\langle \Phi \rangle_k = v_k \to 0$, corresponding to gauge symmetry restoration.

A subtle but important point is that the graviton creation operator is dressed as $\hat{\alpha}^\dagger \sim \hat{N} \hat{a}^\dagger \hat{a}^\dagger$, so that $\hat{\alpha}^\dagger |N\rangle \sim N^2 |N+2\rangle$ for large $N$. This self-referentiality indicates that graviton creation is a collective process whose amplitude depends on the background occupation number, reflecting the Bose-Einstein enhancement and nonlinearities inherent to the graviton condensate. A detailed treatment of the quantum operator algebra lies beyond the present scope. Nevertheless, despite the $k$-dependent scaling of the background metric, the linearized metric perturbations $h_{\mu\nu}^{(k)} = \delta g_{\mu\nu}^{(k)}$ satisfy the standard wave equation $\Box h_{\mu\nu} = 0$ in vacuum, ensuring compatibility with ordinary gravitational wave propagation.

This structure suggests a dual microscopic interpretation of the de Sitter entropy as counting the number of distinct $k$-gravitons corresponding to the same spacetime curvature, with Cosmological Constant $\Lambda_k \sim M_{Pl}/\sqrt{k}$. Thus, the $N$-portrait picture of cosmology postulated by Dvali \& Gomez \cite{Dvali:2012,Dvali:2014,Dvali:2016} emerges naturally from our pre-geometric gauge theory in the Higgsed phase.

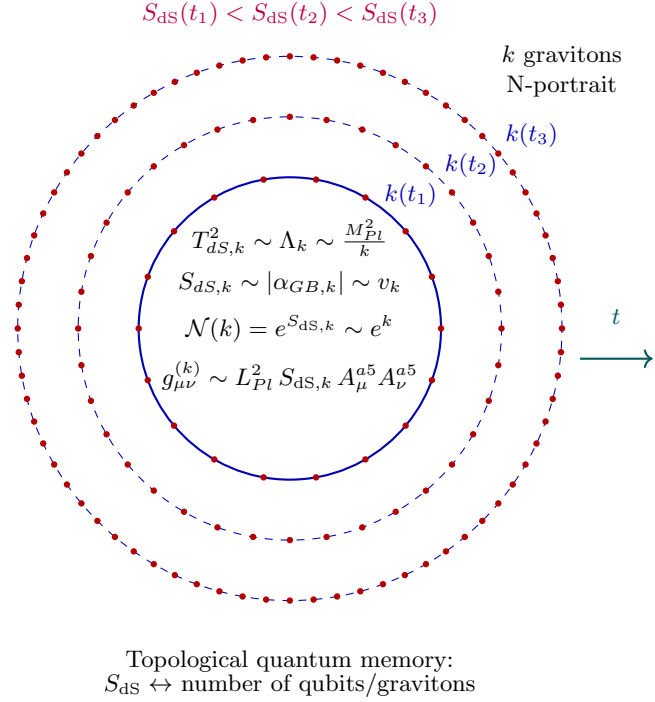
\begin{figure}[t]
\centering
\begin{tikzpicture}[scale=0.8]

\draw[thick, blue!70!black] (0,0) circle (2.5cm);
\node[blue!70!black, font=\small] at (2.0, 2.2) {$k(t_1)$};

\draw[dashed, blue!70!black] (0,0) circle (3.5cm);
\node[blue!70!black, font=\small] at (3.0, 2.7) {$k(t_2)$};

\draw[dashed, blue!70!black] (0,0) circle (4.5cm);
\node[blue!70!black, font=\small] at (4.0, 3.2) {$k(t_3)$};

\foreach \angle in {0,20,40,60,80,100,120,140,160,180,200,220,240,260,280,300,320,340} {
    \pgfmathsetmacro{\x}{2.5*cos(\angle)}
    \pgfmathsetmacro{\y}{2.5*sin(\angle)}
    \fill[red!70!black] (\x,\y) circle (1.5pt);
}

\foreach \angle in {0,10,20,30,40,50,60,70,80,90,100,110,120,130,140,150,160,170,180,190,200,210,220,230,240,250,260,270,280,290,300,310,320,330,340,350} {
    \pgfmathsetmacro{\x}{3.5*cos(\angle)}
    \pgfmathsetmacro{\y}{3.5*sin(\angle)}
    \fill[red!70!black] (\x,\y) circle (1.5pt);
}

\foreach \angle in {0,5,10,15,20,25,30,35,40,45,50,55,60,65,70,75,80,85,90,95,100,105,110,115,120,125,130,135,140,145,150,155,160,165,170,175,180,185,190,195,200,205,210,215,220,225,230,235,240,245,250,255,260,265,270,275,280,285,290,295,300,305,310,315,320,325,330,335,340,345,350,355} {
    \pgfmathsetmacro{\x}{4.5*cos(\angle)}
    \pgfmathsetmacro{\y}{4.5*sin(\angle)}
    \fill[red!70!black] (\x,\y) circle (1.5pt);
}

\node[font=\small, align=center] at (0,1.5) {$T_{dS,k}^2\sim \Lambda_{k}\sim \frac{M_{Pl}^2}{k}$};
\node[font=\small, align=center] at (0,0.75) {$S_{dS,k}\sim |\alpha_{GB,k}|\sim v_{k}$};
\node[font=\small, align=center] at (0,0) {$\mathcal{N}(k)=e^{S_{\text{dS},k}}\sim e^{k}$};

\draw[->, thick, teal!70!black] (4.8, -0.5) -- (6.0, -0.5);
\node[teal!70!black, font=\small] at (5.4, 0.2) {$t$};

\node[purple, font=\small, align=center] at (0, 5.2) {$S_{\text{dS}}(t_1) < S_{\text{dS}}(t_2) < S_{\text{dS}}(t_3)$};

\node[font=\small, align=center] at (4.5, 4.5) {$k$ gravitons};
\node[font=\small, align=center] at (4.5, 4.0) {N-portrait};

\node[font=\small, align=center] at (0, -0.8) {$g_{\mu\nu}^{(k)} \sim L_{Pl}^2\, S_{\text{dS},k}\, A_\mu^{a5}A_\nu^{a5}$};

\node[font=\small, align=center] at (0, -5.5) {Topological quantum memory:};
\node[font=\small, align=center] at (0, -5.9) {$S_{\text{dS}} \leftrightarrow$ number of qubits/gravitons};

\end{tikzpicture}
\caption{\textbf{Holographic interpretation and cosmological evolution}. As the system evolves in time, the de Sitter horizon (blue circle) expands from $k(t_1)$ to $k(t_2)$ to $k(t_3)$ to larger shells (dashed circles), with $k(t_1) < k(t_2) < k(t_3)$. Each shell encodes an increasing number of qubits $\mathcal{N}(k)\sim e^k$, corresponding to growing entropy $S_{\text{dS}}\sim k$ -- with  k displayed as red dots holographically stored on the dS horizon.
The entropy is also scaling as the Gauss-Bonnet topological charge $\alpha_{GB,k}$, the $\theta$-angle and the Higgs VeV $v_{k}$.  The cosmological constant decreases as $\Lambda_k \sim M_{Pl}^2/k$. The N-portrait picture emerges naturally and dS space-time interpreted as a Bose-Einstein condensate of gravitons with Compton wavelength scaling as the Hubble radius. }
\label{fig:holographic}
\end{figure}

\subsubsection{Master Equation and Vacuum Relaxation Dynamics}

We now develop the general dynamical framework describing the quantum evolution of the vacuum energy across the discrete landscape of topological sectors. The key insight is that the total transition rate must include not only the Euclidean bounce action but also the entropy of the final state, which counts the number of available microstates.

For a single transition from sector \(k\) to sector \(k'\), the bare quantum tunneling rate into a \textit{specific} final vacuum state is given by the standard instanton result:
\begin{equation}
\Gamma_{\text{bare}}^{(k \to k')} \sim e^{-\mathcal{S}_E^{(kk')}}
\end{equation}

where \(\mathcal{S}_{E}^{(kk')}\) is the Euclidean bounce action. However, in a landscape with multiple vacua, the \textit{total} rate into \textit{any} state at level \(k'\) must include the degeneracy factor \(\mathcal{N}(k')\), which counts the number of available final states:
\begin{equation}
\Gamma_{k \to k'} = \mathcal{N}(k') \, e^{-\mathcal{S}_E^{(kk')}}
\label{eq:total_rate_general}
\end{equation}

From Eq.~\eqref{eq:dS_quantized}, the de Sitter entropy is quantized.

The number of microstates is related to entropy by the Boltzmann relation:
\begin{equation}
\mathcal{N}(k) = e^{S_{\text{dS}}^{(k)}} \sim e^{k}\, .
\label{eq:microstates_general}
\end{equation}

Therefore, the complete transition rate becomes:

\begin{equation}
\Gamma_{k \to k'} \sim e^{k' - \mathcal{S}_E^{(kk')}}
\label{eq:complete_rate_general}
\end{equation}

Note that the initial sector \(k\) does not receive an entropy factor \(\mathcal{N}(k)\), because the coarse-grained probability \(P_k\) already sums over all microstates in sector \(k\). The rate \(\Gamma_{k \to k'}\) is the transition rate \textit{from} a specific initial microstate (or averaged over initial states with equal weight) \textit{to} any final microstate in sector \(k'\). 
This directly implies that 
\begin{equation}
\Gamma_{k \to k'} \neq \Gamma_{k' \to k}\, ,
\label{eq:rate_asymmetry}
\end{equation}
even in cases where $\mathcal{S}_E^{(kk')}=\mathcal{S}_E^{(kk')}$ -- which also is in general not true.

The probability distribution \(P_k(t)\) over the discrete vacuum sectors evolves according to a Markovian Master Equation:

\begin{equation}
\frac{dP_k}{dt} = \sum_{k' \neq k} \left[ \Gamma_{k' \to k} P_{k'} - \Gamma_{k \to k'} P_k \right]
\label{eq:master_general}
\end{equation}

Here $P_k(t)$ denotes the coarse-grained probability that the Universe resides in vacuum sector \(k\) at time \(t\), marginalized over the $\mathcal{N}(k) \sim e^k$ microstates within that sector and satisfying $\sum_k P_k(t) = 1$.

Using Eq.~\eqref{eq:complete_rate_general}, we obtain:

\begin{equation}
\frac{dP_k}{dt} = \sum_{k' \neq k} \left[ e^{k - \mathcal{S}_E^{(k'k)}} P_{k'} - e^{k' - \mathcal{S}_E^{(kk')}} P_k \right]
\label{eq:master_explicit_general}
\end{equation}

The first term represents transitions \textit{into} sector \(k\) from all other sectors \(k'\), while the second term represents transitions \textit{out of} sector \(k\) to all other sectors.

The Master Equation exhibits a fundamental asymmetry. Consider the ratio of transition rates for upward versus downward adjacent transitions:

\begin{equation}
r_k=\frac{\Gamma_{k \to k+1}}{\Gamma_{k \to k-1}} = \frac{e^{(k+1) - \mathcal{S}_E^{(k,k+1)}}}{e^{(k-1) - \mathcal{S}_E^{(k,k-1)}}} \, . 
\label{eq:asymmetry_general}
\end{equation}
In particular, in any cases where the entropic factor dominates over the Bounce action,
 this ratio is nearly independent of the bounce action \(\mathcal{S}_E^{(k)}\) and 
we obtain $r_k\rightarrow e^2>1$. 
 In these cases, transitions to higher \(k\) (smaller Cosmological Constant) are always entropically favored over transitions to lower \(k\) by a factor of \(e^2\), purely due to the larger number of available states at higher \(k\).
For long jumps, the asymmetry is even more pronounced:
\begin{equation}
\frac{\Gamma_{k \to k+\Delta}}{\Gamma_{k \to k-\Delta}} \sim e^{2\Delta}\, .
\label{eq:asymmetry_long_general}
\end{equation}

This demonstrates that in entropic dominant regime, typical for large k-limit, the system exhibits a \textit{net drift} towards larger values of the topological index \(k\), corresponding to smaller values of the cosmological constant.

The typical rate of change of \(k\) is given by the net transition rate:
\begin{equation}
\frac{dk}{dt} \sim \Gamma_{k \to k+1} - \Gamma_{k \to k-1} \sim e^{k+1 - \mathcal{S}_E^{(k,k+1)}} - e^{k-1 - \mathcal{S}_E^{(k,k-1)}} \, .
\label{eq:dkdt_general}
\end{equation}

Therefore, in large k-limit, we can approximate it as 
\begin{equation}
\frac{dk}{dt} \sim e^{k - \mathcal{S}_E^{(k)}}\, ,
\label{eq:dkdt_general2}
\end{equation}
with $\mathcal{S}_E^{k,k+1}\simeq \mathcal{S}_E^{k,k-1} = \mathcal{S}_E^{k}$ (local micro-reversibility approximation). 

This is the central equation governing the relaxation dynamics. The behavior of the system depends critically on the scaling of \(S_E^{(k)}\) with \(k\).

We now apply the general formalism to the specific scaling predicted by the pre-geometric gravity framework. From Eq.~\eqref{murho}, the scalaron mass scales as:
\begin{equation}
m_\rho^{(k)} \sim M_{Pl} \, k^{-p}
\label{eq:mass_scaling_general}
\end{equation}
where \(p = c_{ssb} + c_{MM} - 3\) encodes the scaling of the SSB parameters. With \(g_{MM} \sim M_{Pl}^{-1}\), the Euclidean action scales as:
\begin{equation}
\mathcal{S}_E^{(k)} \sim g_{MM}^{-2} m_\rho^{-2} \sim k^{2p}
\label{eq:SE_scaling_general}
\end{equation}

As argued above, possible condition in a high-energy SSB framework is that the scalaron mass is \(k\)-independent:
\begin{equation}
m_\rho^{(k)} \sim M_{Pl} \quad \Longrightarrow \quad p = 0\, . 
\label{eq:natural_mass}
\end{equation}
This corresponds to:
\begin{equation}
\mathcal{S}_E^{(k)} = \bar{\mathcal{S}}_{E}
\label{eq:natural_SE}
\end{equation}
with $\bar{\mathcal{S}}_{E}={\rm const}$.

The master equation (\ref{eq:master_explicit_general}) is derived in the WKB (semiclassical) approximation, where the transition rate is dominated by the Euclidean bounce action $\Gamma \sim e^{-\mathcal{S}_E^{(k)}}$. This approximation is valid when the bounce action is large compared to the de Sitter entropy.
\begin{equation}
\mathcal{S}_E^{(k)} \gtrsim S_{dS}^{(k)} \sim k\, .
\label{eq:WKB_validity}
\end{equation}
In this \emph{bounce-dominated regime} but same scaling of the dS-entropy in $k$, the transition rate is suppressed as $e^{-k}$, and the master equation reduces to
\begin{align}
\frac{dk}{dt} &\sim A\,e^{-k}
\quad \Longrightarrow \quad
e^{k}\,dk \sim A\,dt \nonumber \\[4pt]
&\Longrightarrow \quad
k(t) \sim \ln\!\big(A t + e^{k_0}\big) \sim \ln(At)
\label{eq:dkdt_natural}
\end{align}
for \(t \gg t_{Pl}\) and $A=M_{Pl}$. This yields the relaxation law:
\begin{equation}
\boxed{\Lambda(t) \sim \frac{M_{Pl}^2}{\ln(M_{Pl}t)}}\,.
\label{eq:Lambda_natural}
\end{equation}
This case corresponds to 
 \(p =1/2\), i.e., \(m_\rho \sim M_{Pl}/\sqrt{k}\sim \sqrt{\Lambda_{k}}\),
representing the case of a scalaron with the minimal possible mass 
saturating the Hubble rate. Note that the numerical $O(1)$ coefficient multiplying the coupling constant $g_{MM}$ in $\mathcal{S}_{B}$ is a free parameter that determines which contribution dominates---the bounce action or the entropy---when both share the same $k$-scaling.

In the critical case $S_{dS}=\mathcal{S}_{B}$ 
the two factor in the exponential elides 
and the relaxation dynamics can be much faster as
\begin{equation}
\label{Lambdat}
{\it Critical}\,{\it Case}:\quad  \boxed{\Lambda(t) \sim \frac{M_{Pl}}{t+t_{1}}}\,. 
\end{equation}

For generic \(p > 1/2\), the transition rate is
\begin{equation}
\Gamma \sim e^{-\mathcal{S}_E^{(k)} + S_{dS}^{(k)}}
= e^{-k^{2p} + k} \simeq e^{-k^{2p}},
\label{eq:Gamma_generic}
\end{equation}
so the master equation becomes
\begin{equation}
\frac{dk}{dt} \sim A\,e^{-k^{2p}}
\quad \Longrightarrow \quad
\int^{k} e^{u^{2p}}\,du \sim A t .
\label{eq:dkdt_generic}
\end{equation}
For \(2p > 1\), the integral is dominated by the upper limit,
leading to 
\begin{equation}
e^{k^{2p}} \sim A t
\quad \Longrightarrow \quad
k(t) \sim \big(\ln(M_{Pl}t)\big)^{1/(2p)} ,
\label{eq:k_generic}
\end{equation}
and therefore
\begin{equation}
\Lambda(t) \sim \frac{M_{Pl}^2}{\big(\ln(M_{Pl}t)\big)^{1/(2p)}}\, .
\label{eq:Lambda_generic}
\end{equation}
For \(p = 1/2\), this reduces to the logarithmic relaxation
\(\Lambda(t) \sim M_{Pl}^2/\ln(M_{Pl}t)\), while for \(p > 1/2\) the
relaxation is slower, with \(\Lambda(t)\) decaying as an inverse power
of \(\ln(M_{Pl}t)\). 
However, let us note that $p>1/2$ corresponds to 
$\rho$-masses which are below the Hubble rate $H\sim M_{Pl}/\sqrt{k}$,
therefore effectively massless. Thus, we discard this possibility 
since this would introduce massless dilaton like new long-range interactions.

The case \(p < 1/2\) corresponds instead to the regime of entropic
domination, where the barrier is completely negligible compared to the
entropic factor. The evolution is then entirely thermal, independent of
the quantum tunneling rate.
This also include the case $p=0$ which corresponds to a k-scale invariant 
Euclidean action $S_{E}=\bar{S}_{E}={\rm const}$
and $m_{\rho}\sim M_{Pl}$. 

\begin{figure}[t]
\centering
\begin{tikzpicture}[scale=0.9]
\draw[->] (0,0) -- (5.5,0) node[right] {$k(t)$};
\draw[->] (0,0) -- (0,5) node[above] {$\epsilon_k^2 / \epsilon_{1}^2$};

\fill[gray!20] (0,0) rectangle (1,5);
\node[rotate=90, gray!60!black] at (0.5, 2.5) {PGG phase};

\foreach \k in {1,2,3,4,5} {
    \pgfmathsetmacro{\y}{5/\k}
    \fill (\k,\y) circle (2pt);
    \draw[dashed, gray] (\k,0) -- (\k,\y);
}

\draw[domain=1:5.2, smooth, red, dashed] plot (\x, {5/\x});

\draw[->, red, thick] (5.2, 0.4) -- (7.5, 0.4);
\node[red] at (7.0, 0.85) {$t \rightarrow  \infty$, $\Lambda_{\infty}=0$};

\node at (1, -0.3) {$1$};
\node at (2, -0.3) {$2$};
\node at (3, -0.3) {$3$};
\node at (4, -0.3) {$4$};
\node at (5, -0.3) {$5$};

\node[red] at (3.5, 3.0) {$\Lambda_k(t) \sim M_{Pl}^2/k(t)$};

\draw[dotted, thick] (1,0) -- (1,5);
\end{tikzpicture}
\caption{Discrete spectrum of vacuum energy levels in pre-geometric gravity. The vertical axis shows the normalized vacuum energy $\epsilon_k^2/\epsilon_1^2 \sim 1/k$, while the horizontal axis represents the topological index $k(t)$, which increases with cosmological time. The shaded region corresponds to the pre-geometric (PGG) phase where the semi-classical description breaks down. The discrete points (filled circles) represent the quantized vacuum states for integer $k \geq 1$, following the asymptotic scaling $\Lambda_k \sim M_{Pl}^2/k$ (red dashed curve). The system evolves toward larger $k$ (smaller $\Lambda$), asymptotically approaching the attractor $\Lambda_\infty = 0$ as $t \to \infty$. The vertical dotted line at $k=1$ marks the boundary between the pre-geometric and geometric phases.}
\label{fig:spectrum}
\end{figure}
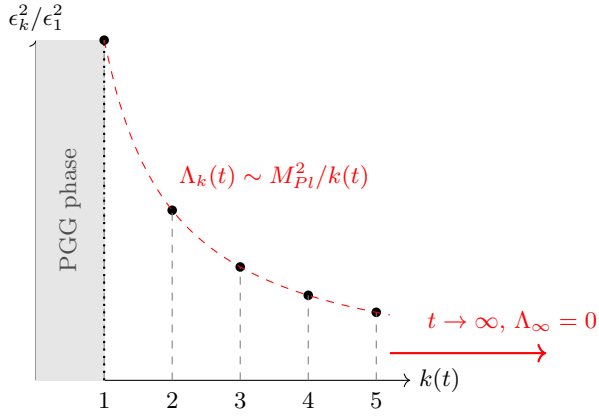

In all cases $p\geq 1/2$, the system flows toward the asymptotic attractor:
\begin{equation}
\boxed{\lim_{t \to \infty} k(t) = \infty \quad \Longrightarrow \quad \lim_{t \to \infty} \Lambda(t) = 0}
\label{eq:attractor_general}
\end{equation}

This represents a universal feature of $p\geq 1/2$ case: \(\Lambda_\infty = 0\) is the unique attractor of the relaxation dynamics. However, the relaxation time diverges as \(\bar{t} \to \infty\) in this limit. The observed small but nonzero value of the cosmological constant may therefore be understood as a reflection of the finite age of the Universe: we occupy the sector \(k \sim 10^{120}\) simply because the Universe has had a finite time to relax from its initial state.

For \(p \leq 1/2\) the situation is more subtle. Considering the
entropy-dominated equation
\begin{equation}
\label{entropyDom}
\frac{dk}{dt}\sim A\,e^{k}
\quad \Longrightarrow \quad
k(t) \sim -\ln\!\big(e^{-k_0} - A t\big),
\end{equation}
which diverges at the finite time \(t_* = e^{-k_0}/A\). This
finite-time singularity signals a breakdown of the effective
description since an infinite tower of higher order terms 
has to be included. More than a relaxation, this suggests an uncontrolled cascade toward 
a vacuum energy density decay. 
Note that in the case $A=M_{Pl}$ and $k_{0}\sim 1$ this process will be in a planckian time!
We postpone more analysis of this scenario 
for future works. 


A brief comment on the interplay between quantum break-time and vacuum relaxation is in order. In the cases $p\geq 1/2$, and in particular the most physically motivated case $p=1/2$, the quantum break-time $t_{\mathrm{q}} \sim t_{Pl} k^{3/2}$ is parametrically shorter than the vacuum tunneling time $t_{\text{relax}} \sim t_{Pl} e^k$ $(p=1/2)$. However, this does not obstruct the relaxation; rather, the decoherence processes responsible for the quantum break-time provide an alternative channel for vacuum transitions. Individual coherent-state gravitons can scatter out of the condensate more efficiently than the full instanton transition, effectively accelerating the escape rate from
\begin{equation}
\Gamma_{\text{tunnel}} \sim e^{-k} \quad \longrightarrow \quad \Gamma_{\text{escape}} \sim \frac{1}{t_{\mathrm{q}}} \sim t_{Pl}^{-1} k^{-3/2}.
\end{equation}
Thus, the quantum break-time enhances the barrier tunneling from an exponential suppression to a power-law process, as coherent quanta can more easily tunnel out of the condensate while preserving a given vacuum sector $v_k$. In the critical case $m_\rho \sim M_{Pl}/\sqrt{k}$, the relaxation time $t_{\text{relax}} \sim t_{Pl} k$ is already faster than the quantum break-time $t_{\mathrm{q}} \sim t_{Pl} k^{3/2}$, so the system tunnels to smaller $\Lambda$ sectors before decoherence becomes relevant. Consequently, in both cases the system flows toward the asymptotic attractor $\Lambda_\infty = 0$, albeit through different dynamical regimes.

These considerations invite speculation on the possibility of a pre-Big Bang phase, whose memory might be encoded in the present smallness of the Cosmological Constant. A dynamical decreasing \(\Lambda\) suggests that in the far future the cosmic acceleration will gradually diminish, potentially leading to an expansion turning point and subsequent collapse—a scenario reminiscent of bounce or cyclic cosmologies. Within the pre-geometric gravity framework, such a critical bounce could be interpreted as a gauge and topological phase transition. Quantum corrections may further modify the equation of state of dark energy, accelerating the asymptotic relaxation. It is also conceivable that the observed dark energy is not the tail of the asymptotic relaxation but rather a temporary metastable state in a more complex landscape. We leave a detailed exploration of these possibilities for future work.

The entropic asymmetry in the transition rates provides a natural explanation for the arrow of time. The Master Equation \eqref{eq:master_explicit_general} is manifestly irreversible: the system evolves toward higher \(k\) (larger entropy) and cannot spontaneously return to lower \(k\) without violating the second law of thermodynamics. The probability of a downward transition from \(k\) to \(k-\Delta\) is suppressed by \(e^{-2\Delta}\), as follows from Eq.~\eqref{eq:asymmetry_long_general}. For a transition from the observed sector \(k \sim 10^{120}\) down to the Planckian sector \(k \sim 1\), the suppression is
\begin{equation}
\Gamma_{k \to 1} \sim e^{-2 \times 10^{120}},
\end{equation}
which is effectively zero for all practical purposes. This ensures that the cosmological relaxation is a one-way process, with the Universe evolving toward ever smaller cosmological constant and asymptotically approaching \(\Lambda_\infty = 0\).

This thermodynamic arrow is not imposed externally but emerges dynamically from the topological structure of the pre-geometric gauge theory, providing a fundamental connection between quantum gravity, entropy, and the observed arrow of time. The entropy factor \(\mathcal{N}(k) \sim e^k\) is precisely the holographic counting of degrees of freedom on the de Sitter horizon. The de Sitter entropy \(S_{\text{dS}}^{(k)} \sim k\) serves simultaneously as the topological index and the measure of available microstates. This provides a concrete realization of the Holographic Naturalness proposal \cite{HN1,HN2,HN3,Addazi:2026szh}: the observed smallness of the Cosmological Constant is protected by the enormous entropy of the horizon, which makes transitions to lower entropy states exponentially improbable. The vacuum energy relaxation described here is therefore not a fine-tuning but a thermodynamic inevitability: the Universe simply evolves toward its most probable state, which is the one with maximal entropy and minimal cosmological constant.

From a microscopic perspective, the increase in \(k\) corresponds to an increasing number of soft graviton states, which can be understood as an ultra-cold condensate. In the thermodynamic picture, the cosmological constant sets a de Sitter temperature \(T \sim \sqrt{\Lambda}\); the relaxation dynamics thus corresponds to a spontaneous cooling of the graviton system. This may be interpreted as a splitting dynamics, such as graviton rescatterings \(2 \to m\) with \(m > 2\), which dynamically increase the number of states with lower-energy single partons emerging from the collision center-of-mass energy. In this sense, the arrow of time can emerge microscopically as an overall macroscopic effect of high-multiplicity graviton splitting scatterings. Indeed, in our picture the metric tensor and gravitons are emergent condensates rather than fundamental fields, and as quantum operators they contain an extra quantum number operator compared to the traditional quantum gravity picture. This suggests that graviton creation operators increase with the graviton occupation number, leading to spontaneous multiplicity growth. As shown in Ref.~\cite{Addazi:2025zqf}, the insertion of such soft gravitons with Standard Model propagating fields can cure UV divergences coming from virtual bubble diagrams, effectively replacing \(M_{Pl}^2\) with \(\Lambda_k\).

\subsubsection{On the Possible Implications for DESI  }

Given that $\Lambda_k \sim 1/k$ evolves with cosmic time---corresponding to the critical evolution case---we have
\begin{equation}
\dot{\Lambda}_k/\Lambda_k = -\dot{k}/k \,,
\end{equation}
and for the dark energy (DE) component
\begin{equation}
\dot{\rho}_{\rm DE} + 3H(1+w)\rho_{\rm DE} = 0 \,, \qquad \dot{k} = 3Hk(1 + w) \,.
\end{equation}
A natural parametrization is provided by $w_0 w_a$CDM:
\begin{equation}
w(z) = w_0 + w_a \frac{z}{1+z} \,,
\end{equation}
which allows the transition rate and other quantities to be estimated. A dedicated comparison with DESI---especially given the reported $2.8$--$4.2\sigma$ preference for a ``phantom crossing'' equation of state ($w_0 > -1$, $w_a < 0$, crossing at $z \sim 0.5$) \cite{DESI:ExtendedDE, Yang:2025QuintomMG, Cai:2026QuintomReview}---remains an important direction for future work.

Furthermore, recent studies have shown that higher curvature corrections to the pre-metric MM action predict a phantom-crossing dynamics capable of explaining the DESI anomaly \cite{Addazi:2026ulq}. The interplay between higher curvature corrections and the multi-quantum tunneling dynamics considered in this work could genuinely open up a whole phenomenology in dynamical dark energy, whose full exploration warrants more efforts beyond the scope of this work.

\vspace{0.1cm}

\subsection{Conclusions}

We have investigated a quantum relaxation mechanism for the vacuum energy density within MacDowell--Mansouri pre-geometric gravity. The spontaneous breaking $SO(1,4)\to SO(1,3)$ leads to a countably infinite set of vacua labeled by a topological index $k$, arising from the periodicity of the Gauss-Bonnet angle and gravitational Pontryagin angle. These vacua are separated by potential barriers, allowing for quantum tunneling transitions.

Our central result is that the system preferentially evolves along the ordered sequence $ \Lambda_1 > \cdots > \Lambda_k > \cdots$, where the Cosmological Constant gradually decreases  toward the asymptotic attractor $\Lambda_\infty = 0$. This behavior is driven by the entropic favorability of higher topological sectors, with the number of available microstates growing as $\mathcal{N}(k) \sim e^k$, directly related to the quantized de Sitter entropy $S_{\text{dS}}^{(k)} \sim k$. The resulting Master Equation exhibits fundamental irreversibility, providing a possible explanation for the arrow of time from first principles.

The quantization of vacuum energy, $\Lambda_k \sim M_{Pl}^2/k$, offers a new perspective on the Cosmological Constant problem. In this framework, the observed value $\Lambda \sim 10^{-120} M_{Pl}^2$ may be understood not as a fine-tuned parameter but as a reflection of the finite age of the Universe: we occupy the sector $k \sim 10^{120}$ because the Universe has had finite time to relax from an initial Planckian vacuum.

The framework suggests a re-interpretation of de Sitter entropy as a topological quantum memory, with the graviton emerging as a collective excitation of the underlying condensate, naturally reproducing aspects of the N-portrait picture \cite{Dvali:2012,Dvali:2014,Dvali:2016}. Extended versions of the model may offer testable predictions through dynamical dark energy~\cite{Addazi:2026ulq} as a possible explanation of DESI data \cite{DESI:ExtendedDE, Yang:2025QuintomMG, Cai:2026QuintomReview}.
Crucially, the relaxation dynamics derived here is not a phenomenological ansatz but a direct consequence of the pre-geometric gauge framework. The countable infinity of vacua, the discrete spectrum $\Lambda_k \sim M_{Pl}^2/k$ and the entropic irreversibility of the Master Equation,  all emerge from the Spontaneous Symmetry Breaking  $SO(1,4) \to SO(1,3)$ and the periodicity of the gravitational $\theta$-angle. If realized in nature, this framework would resolve the Cosmological Constant problem dynamically: entropic selection drives the system toward smaller $\Lambda$, while the arrow of time emerges from the irreversibility of vacuum relaxation.

\vspace{0.2cm}

{\bf Acknowledgments}. 
We thank the anonymous referee for the valuable comments
and remarks, which have significantly improved our manuscript.
My work was supported by Program for Innovative Research Team in Anqing Normal University and the National Science Foundation of China (NSFC) through the grant No. 12350410358. 
We also acknowledge that DeepSeek 3.0 was used to supervise linguistic checks.

\vspace{0.1cm}
{\it Declarations}. The author declares no conflict of interests.

{\it Data Availability Statement}: No Data associated in the manuscript.


\begin{thebibliography}{99}

\bibitem{Addazi:2024rzo}
A.~Addazi, S.~Capozziello, A.~Marciano and G.~Meluccio,
Class. Quant. Grav. \textbf{42} (2025) no.4, 045012
doi:10.1088/1361-6382/ada767
[arXiv:2409.02200 [hep-th]].

\bibitem{Addazi:2025vbw}
A.~Addazi, S.~Capozziello, A.~Marcian{\`o} and G.~Meluccio,
Phys. Rev. D \textbf{112} (2025) no.6, 064058
doi:10.1103/v5wg-sy4w
[arXiv:2505.01272 [gr-qc]].

\bibitem{Addazi:2025gcl}
A.~Addazi,
[arXiv:2512.20681 [physics.gen-ph]].

\bibitem{Addazi:2025zqf}
A.~Addazi, S.~Capozziello and G.~Meluccio,
MDPI Physics \textbf{8} (2026) no.1, 2
doi:10.3390/physics8010002
[arXiv:2604.26032 [hep-th]].

\bibitem{Addazi:2025qkc}
A.~Addazi, S.~Capozziello, J.~Liu, A.~Marciano, G.~Meluccio and X.~L.~Su,
[arXiv:2505.17014 [gr-qc]], 


{
\bibitem{Addazi:2026suu}
A.~Addazi and G.~Meluccio,
Phys. Rev. D \textbf{114} (2026) no.4, 044056
doi:10.1103/fytf-m83r
[arXiv:2609.10619 [gr-qc]].

\bibitem{Addazi:2026eto}
A.~Addazi,
Eur. Phys. J. C \textbf{86} (2026) no.8, 916
doi:10.1140/epjc/s10052-026-16181-2}




\bibitem{MacDowell77} S. W. MacDowell and F. Mansouri, \textit{Phys. Rev. Lett.} \textbf{38}, 739 (1977).

\bibitem{SA1}
    A. H. Chamseddine, P. C. West, Supergravity as a gauge theory of supersymmetry, Nucl. Phys. B \textbf{129}, 39--44 (1977).
\bibitem{SA2}
    K. S. Stelle, P. C. West, de Sitter gauge invariance and the geometry of the Einstein-Cartan theory, J. Phys. A Math. Gen. \textbf{12}, L205 (1979).

\bibitem{Wilczek98}
F. Wilczek, 
Phys. Rev. Lett. 80, 4851-4854 (1998).


\bibitem{Weinberg89} S. Weinberg, \textit{Rev. Mod. Phys.} \textbf{61}, 1 (1989).


\bibitem{Immirzi97} G. Immirzi, \textit{Class. Quant. Grav.} \textbf{14}, L177 (1997).

{\bibitem{NY}
    H. T. Nieh, M. L. Yan, 
    J. Math. Phys. \textbf{23}, 373--374 (1982).}

\bibitem{Holst96} S. Holst, \textit{Phys. Rev. D} \textbf{53}, 5966 (1996).

\bibitem{Rovelli2004}
C. Rovelli, \textit{Quantum Gravity} (Cambridge University Press, 2004).


\bibitem{CSK1}
H. Kodama, 
Phys. Rev. D {\bf 42}, 2548 (1990).

\bibitem{CSK2}
L. Freidel and L. Smolin,
Class. Quant. Grav. {\bf 21}, 3831 (2004),
arXiv:hep-th/0310224.


\bibitem{Alexander:2022ocp}
S.~Alexander, G.~Herczeg and L.~Freidel,
Class. Quant. Grav. \textbf{40} (2023) no.14, 145010
doi:10.1088/1361-6382/acde3d
[arXiv:2212.07446 [hep-th]].

\bibitem{Magueijo:2020ugp}
J.~Magueijo,
Phys. Rev. D \textbf{102} (2020) no.4, 044034
doi:10.1103/PhysRevD.102.044034
[arXiv:2005.03381 [gr-qc]].


\bibitem{Alexander:2025qkx}
S.~Alexander, H.~Bernardo and A.~Hui,
Phys. Rev. Lett. \textbf{136} (2026) no.15, 151501
doi:10.1103/rzz5-p4f4
[arXiv:2506.14886 [gr-qc]].

{\bibitem{Coleman1977}
S.~Coleman,
Phys.\ Rev.\ D \textbf{15}, 2929 (1977).

\bibitem{Coleman1985}
S.~Coleman,
``The uses of instantons,''
in \textit{Aspects of Symmetry}
(Cambridge University Press, 1985).

\bibitem{Linde1983}
A.~D. Linde,
Nucl.\ Phys.\ B \textbf{216}, 421 (1983).}


\bibitem{GB}
G. W. Gibbons and S. W. Hawking, 
Phys. Rev. D {\bf 15}, 2738-2751 (1977) doi:10.1103/PhysRevD.15.2738

\bibitem{Dvali:2012}
G.~Dvali and C.~Gomez,
Fortsch. Phys. \textbf{61}, 742-767 (2013)
doi:10.1002/prop.201300001
[arXiv:1112.3359 [hep-th]].

\bibitem{Dvali:2014}
G.~Dvali and C.~Gomez,
JCAP \textbf{1401}, 023 (2014)
doi:10.1088/1475-7516/2014/01/023
[arXiv:1312.4795 [hep-th]].

\bibitem{Dvali:2016}
G.~Dvali and C.~Gomez,
Annalen Phys. \textbf{528}, 68-73 (2016)
doi:10.1002/andp.201500216
[arXiv:1412.8077 [hep-th]].


\bibitem{HN1}
A.~Addazi, 
 Int. J. Mod. Phys. D 2020, {\bf 29}, 2050084. https://doi.org/10.1142/ S0218271820500844.

\bibitem{HN2}
A.~Addazi, 
Eur. Phys. J. Plus 2020, {\bf 135}, 940. https://doi.org/ 10.1140/epjp/s13360-020-00933-4.

\bibitem{HN3}
A.~Addazi, 
Int. J. Geom. Meth. Mod. Phys. 2021, {\bf 18}, 2150030. https://doi.org/10.1142/S0219887821500304.

\bibitem{Addazi:2026szh}
A.~Addazi and G.~Meluccio,
Particles \textbf{9} (2026) no.1, 11
doi:10.3390/particles9010011
[arXiv:2604.26982 [hep-ph]].

{\bibitem{DESI:ExtendedDE}
K.~Lodha \textit{et al.} [DESI Collaboration],
Phys. Rev. D \textbf{112}, 083511 (2025),
\texttt{arXiv:2503.14743 [astro-ph.CO]}.


\bibitem{Yang:2025QuintomMG}
Y.~Yang, Q.~Wang, X.~Ren, E.~N.~Saridakis and Y.-F.~Cai,
Astrophys. J. \textbf{988}, 123 (2025),
\texttt{arXiv:2504.06784 [astro-ph.CO]}.

\bibitem{Cai:2026QuintomReview}
Y.-F.~Cai, X.~Ren, T.~Qiu, M.~Li and X.~Zhang,
\texttt{arXiv:2505.24732 [astro-ph.CO]}.

\bibitem{Brout:2022PantheonPlus}
D.~Brout \textit{et al.},
Astrophys. J. \textbf{938}, 110 (2022),
\texttt{arXiv:2202.04077 [astro-ph.CO]}.

\bibitem{Rubin:2025Union3}
D.~Rubin \textit{et al.},
Astrophys. J. \textbf{986}, 231 (2025),
\texttt{arXiv:2311.12098 [astro-ph.CO]}.

\bibitem{DES:2024Y5}
T.~M.~C.~Abbott \textit{et al.} [DES Collaboration],
Astrophys. J. Lett. \textbf{973}, L14 (2024),
\texttt{arXiv:2401.02929 [astro-ph.CO]}.

\bibitem{Addazi:2026ulq}
A.~Addazi, Y.~Aldabergenov, D.~Berkimbayev, Y.~Cai, S.~Capozziello and G.~Meluccio,
[arXiv:2605.07344v2 [gr-qc]], to appear in SCPMA journal.}


%
%
%
%




\end{thebibliography}
\end{document}